\documentclass{article}
\usepackage{amssymb}
\usepackage{amsfonts}
\usepackage{amsmath}
\usepackage[dvipdf]{graphicx}

\input{tcilatex}

\begin{document}

\title{Comment on $"$Position-dependent effective mass Dirac equations with $%
PT$- symmetric and non - $PT$- symmetric potentials$"$ [J. Phys. A: Math.
Gen. 39 (2006) 11877--11887]}
\author{Omar Mustafa$^{1}$ and S.Habib Mazharimousavi$^{2}$ \\
%EndAName
Department of Physics, Eastern Mediterranean University, \\
G Magusa, North Cyprus, Mersin 10,Turkey\\
$^{1}$E-mail: omar.mustafa@emu.edu.tr\\
$^{2}$E-mail: habib.mazhari@emu.edu.tr}
\maketitle

\begin{abstract}
Jia and Dutra (J. Phys. A: Math. Gen. 39 (2006) 11877) have considered the
one-dimensional non-Hermitian complexified potentials with real spectra in
the context of position-dependent mass in Dirac equation. In their second
example, a smooth step shape mass distribution is considered and a
non-Hermitian non - $PT$- symmetric Lorentz vector potential is obtained.
They have mapped this problem into an exactly solvable Rosen-Morse Schr\"{o}%
dinger model and claimed that the energy spectrum is real. The energy
spectrum they have reported is pure imaginary or at best forms an empty set.
Their claim on the reality of the energy spectrum is fragile, therefore.

\end{abstract}

Very recently, Jia and Dutra [1] have presented a new procedure to construct
a set of non-Hermitian potentials with real spectra for the one-dimensional
Dirac particle endowed with position-dependent mass. They have reported that
a Lorentz vector potential of the form

\begin{equation}
V(x)=\frac{i}{2}\frac{1}{M(x)}\frac{dM(x)}{dx}
\end{equation}%
would imply a Schr\"{o}dinger-like Dirac equation 
\begin{equation}
-\frac{d^{2}}{dx^{2}}\varphi (x)+V_{eff}(x)\varphi (x)=E^{2}\varphi (x),
\end{equation}%
where $V_{eff}(x)=M(x)^{2}$, $M(x)$ and $E$ are the mass and the real energy
of the Dirac's particle, respectively, and $\varphi (x)$ is the upper
component of the Dirac spinor (eqs. (1)-(9) in [1]).

In their second example (section 3.2 in [1]) they have considered a smooth
step shape mass distribution%
\begin{equation}
M\left( x\right) =M_{\circ }\left( 1+\eta \tanh \alpha x\right)
\end{equation}
that led to a non-$\mathcal{PT}$-symmetric non-Hermitian Lorentz vector
potential%
\begin{equation}
V\left( x\right) =\frac{i}{2}\frac{\alpha \eta \func{sech}^{2}\alpha x}{%
\left( 1+\eta \tanh \alpha x\right) }
\end{equation}%
where $\eta $ is a small parameter that satisfies the condition $\left\vert
\eta \right\vert $ $\leq 1$ to get a positive mass distribution. Under such
settings, Eq.(2) reads a Rosen- Morse-type Schr\"{o}dinger problem [2].
Employing the SUSQM method [3], Jia and Dutra [1] have reported that the
reality of energy spectrum%
\begin{equation}
E_{n}=\pm \sqrt{M_{\circ }^{2}(1+\eta ^{2})-\frac{\eta ^{2}M_{\circ }^{4}}{%
\alpha ^{2}(n+\delta _{1})^{2}}-\alpha ^{2}(n+\delta _{1})^{2}},\text{ \ }%
n=0,1,2,\cdots ,
\end{equation}%
with 
\begin{equation}
\delta _{1}=\frac{1}{2}\left( 1-\sqrt{1+\frac{4\eta ^{2}M_{\circ }^{2}}{%
\alpha ^{2}}}\right) ,
\end{equation}%
is guaranteed if and only if%
\begin{equation}
M_{\circ }^{2}(1+\eta ^{2})\geq \frac{\eta ^{2}M_{\circ }^{4}}{\alpha
^{2}(n+\delta _{1})^{2}}+\alpha ^{2}(n+\delta _{1})^{2}.
\end{equation}

In this comment we shall show that even with the lowest possible value of
the principle quantum number $n$ (i.e., $n=0$) the inequality in (7) can not
be satisfied. Hence, the reality of the spectrum in (5) is fragile (of
course within the parameters' settings reported in [1] for this particular
case).\FRAME{ftFU}{4.0542in}{3.0234in}{0pt}{\Qcb{A 3D plot of $f(\protect%
\eta ,\Lambda )$ with $\Lambda \in \left[ 0,10\right] $ and $-1\leq \protect%
\eta $ $\leq 1$.}}{}{Figure}{\special{language "Scientific Word";type
"GRAPHIC";maintain-aspect-ratio TRUE;display "PICT";valid_file "T";width
4.0542in;height 3.0234in;depth 0pt;original-width 4.0041in;original-height
2.9793in;cropleft "0";croptop "1";cropright "1";cropbottom "0";tempfilename
'../Local Settings/Temp/Dutra.wmf';tempfile-properties "NPR";}}

To do so, we start with the introduction of a positive parameter $\Lambda
=\alpha /M_{\circ }\geq 0$ to recast, with $n=0$, the inequality (7) as 
\begin{equation}
f(\eta ,\Lambda )=(1+\eta ^{2})-\frac{4\eta ^{2}}{\left( \Lambda -\sqrt{%
\Lambda ^{2}+4\eta ^{2}}\right) ^{2}}-\frac{1}{4}\left( \Lambda -\sqrt{%
\Lambda ^{2}+4\eta ^{2}}\right) ^{2}\geq 0.
\end{equation}%
Then we shall now test the validity of $%
%TCIMACRO{\U{211d} }%
%BeginExpansion
\mathbb{R}
%EndExpansion
\ni f(\eta ,\Lambda )\geq 0$ for $-1\leq \eta $ $\leq 1$ and $\Lambda \geq 0$%
, where $\eta ,\Lambda \in 
%TCIMACRO{\U{211d} }%
%BeginExpansion
\mathbb{R}
%EndExpansion
$.

A 3D plot, in Figure 1, of $f(\eta ,\Lambda )$ with $\Lambda \in \left[ 0,10%
\right] $ clearly indicates that $0\geq 
%TCIMACRO{\U{211d} }%
%BeginExpansion
\mathbb{R}
%EndExpansion
\ni f(\eta ,\Lambda )\ngeqslant 0$ over the physically acceptable parametric
values. Therefore, the energy spectrum in (5) can only be pure imaginary
(i.e., $E_{n}\in 
%TCIMACRO{\U{2102} }%
%BeginExpansion
\mathbb{C}
%EndExpansion
$ and $E_{n}\notin 
%TCIMACRO{\U{211d} }%
%BeginExpansion
\mathbb{R}
%EndExpansion
$)

\qquad


\begin{thebibliography}{9}
\bibitem{} Jia C S and de Souza Dutra A, 2006 J. Phys. A: Math. Gen. \textbf{%
39} 11877

\bibitem{} Rosen N and Morse P M, 1932 Phys. Rev. \textbf{42 }210

\bibitem{} F Cooper, A Khare and U Sukhatme, (1995) Phys. Rep. \textbf{251}
267
\end{thebibliography}
\end{document}